\documentclass[
preprint,
 amsmath,amssymb,
 aps,
floatfix,
]{revtex4-2}

\usepackage{comment}
\usepackage{graphicx}
\usepackage{dcolumn}
\usepackage{bm}
\usepackage[version=3]{mhchem}
\usepackage{tabularx}
\usepackage{xcolor} 
\newcolumntype{P}[1]{>{\centering\arraybackslash}p{#1}}

\begin{document}

\title{Unveiling ZIF-8 nucleation mechanisms through molecular simulation: role of temperature, solvent and reactant concentration}

\author{Sahar Andarzi Gargari}
\author{Rocio Semino}%
 \email{rocio.semino@sorbonne-universite.fr}
\affiliation{Sorbonne Université, CNRS, Physico-chimie des Electrolytes et Nanosystèmes Interfaciaux, PHENIX, F-75005 Paris, France.}%

\begin{minipage}\textwidth

\begin{abstract}
Synthesizing new metal-organic frameworks (MOFs) is a challenging task, as the size, morphology, polymorph and type and number of defects present on the synthesis product may depend on many variables, including temperature, solvent, concentration and nature of reactants, among others. A deeper understanding on how synthesis conditions determine the obtained material is crucial to optimize the use of resources when synthesizing new MOFs. In this contribution, we study the impact of changing concentration, solvent and temperature on the molecular level mechanisms of the solvothermal nucleation process of ZIF-8 relying on molecular dynamics simulations using a force field that incorporates metal-ligand reactivity. We find that the nucleation is faster when the synthesis is performed in dimethylsulfoxide than when it is performed in methanol, in alignment with experimental observations. In the early steps of the nucleation process, we observe the formation of linear oligomers containing metal ions and ligands, which start forming cycles later on. All simulations lead to the formation of a final state that is highly-connected and partially amorphous, which could be correlated to an intermediate species observed in direct experiments. The mechanism of formation of this phase mainly consists of the merging of smaller nuclei. Even though increasing temperature and the reactants concentration lead to a similar nucleation speed-up, there are differences in ring populations and lifetimes within the highly-connected amorphous intermediate phases that are formed in each case. Finally, important differences in the free energy of Zn--2-methylimidazolate versus Zn--imidazolate subsequent binding events are revealed and discussed.   
\end{abstract}
\maketitle
\end{minipage}

\section{Introduction}

New materials can help solve some of the most pressing societal issues from our era, including capturing and storing pollutants,\cite{Rojas2020,Jin2023,MartnezAhumada2021, Dutta2021} providing new paths for drug delivery\cite{Osterrieth2020,Bunzen2021} and acting as catalysts\cite{Shen2021,Jin2024,NarvezCelada2022} for greener chemistry. In this context, metal-organic frameworks (MOFs) offer materials scientists a platform to test how changing structural characteristics, including acid/base, hydrophobic/hydrophillic and/or redox behaviour or pore sizes, can affect the properties of the resulting material. Indeed, MOFs can be made with a wide variety of metal-based clusters and organic polidentate ligands, which allows for a huge versatility in the creation of new MOF structures that can be futher tested for target applications. However, despite the important progress made in the reticular chemistry field,\cite{Chen2023,Kalmutzki2018,Jiang2021} our lack of molecular level understanding of the mechanisms that lead from the reactants to the resulting MOF as a function of synthesis conditions hampers our capacity to design MOFs \textit{à-la-carte}. This topic has been sparking the curiosity of many researchers for more than a decade.\cite{Bhawnani2025,Forgan2020,Cheetham2018,Millange2010,Stavitski2011,He2021,Moh2013,Cravillon2012,Talosig2024,Dok2025}

The synthesis process of ZIF-8,\cite{Park2006} a highly stable zeolitic imidazolate framework, has been widely studied by the research community. This MOF is composed by Zn$^{2+}$ ions that form coordination bonds with 2-methylimidazolate (MIm$^-$) ligands following a tetrahedral geometry, and it adopts the sodalite topology. It has been tested for a wide range of processes, including biomedical applications, capture of pollutants and separation of hydrocarbons,\cite{Abdelhamid2021,Nazir2025,Bergaoui2021} and has been synthesized in a wide range of synthesis conditions.\cite{Lee2015} Many solvents, including dimethylformamide (DMF),\cite{Park2006} water,\cite{Pan2011} methanol (MeOH),\cite{Cravillon2012} dimethyl sulfoxide (DMSO)\cite{Feng2016}  and mixtures have been used. Kinetic studies by \textit{in situ} EDXRD have shown that the self-assembly of ZIF-8 is controlled by nucleation when the synthesis is carried out in DMF,\cite{Moh2013} while methanol-based solvothermal synthesis has shown similar rates of nucleation and growth.\cite{Cravillon2012} Bustamante and coworkers have correlated the hydrogen bonding donation capacity of the solvent with nucleation kinetics.\cite{Bustamante2014} The effect of temperature on ZIF-8 methanol-based solvothermal synthesis was investigated in several studies, and while higher temperatures have resulted in smaller particle sizes in some cases\cite{Tsai2016}, other works have shown negligible effect of temperature.\cite{Beh2018} Beh and coworkers have also shown that particle size can be modulated by changing the concentration of the reactants.\cite{Beh2018} Malekmohammadi and collaborators have studied the effect of temperature, ligand to metal ratio and of the presence of salts in ZIF-8 synthesis in methanol/water mixtures.\cite{Malekmohammadi2019} Another synthesis variable that can greatly affect the properties of the generated materials is pH.\cite{Kida2013,Jin2023} Manipulating synthesis conditions can also give rise to the formation of alternative polymorphs for ZIF-8.\cite{Talosig2024,Jin2023,Balog2022} These studies illustrate how synthesis mechanisms are strongly dependent on synthesis conditions and that studying them is not an easy task, as it requires combining an ensemble of techniques to address all the different time- and length-scales involved in this complex chemical reaction. 

Despite the breadth of experimental studies devoted to better understanding the impact of synthesis conditions on the properties of ZIF-8, just a handful simulation studies were only very recently dedicated to this topic, mainly because modeling nucleation is very challenging.\cite{Neha2022,Finney2023,Bhawnani2025} Balestra and Semino have developed nb-ZIF-FF: a partially reactive force field for 2-methylimidazolate and imidazolate containing Zn-based ZIFs, where only the coordination bonds are allowed to dynamically form and break, while all other bonds within the ligands are modeled through harmonic potentials.\cite{Balestra2022} This force field, validated for ZIF-8 among other MOFs, has been developed and applied in our group to model the early stages of nucleation of ZIF-8 in methanol \textit{via} well-tempered metadynamics (WT-MetaD)\cite{Barducci2008} simulations,\cite{Balestra2022} and to model early nucleation and late growth of a series of imidazole-based ZIFs\cite{Mendez2025}. These simulations reveal a disordered but highly-connected intermediate structure during the formation of the material, which resembles intermediate species observed in direct experimental synthesis.\cite{Venna2010,Jin2023,Dok2025, Balog2022,Talosig2024} The formation of building units was partially covered in this work, but a more in-depth investigation, where other solvents were also considered, was made by density functional theory calculations.\cite{Balestra2023} An in-depth combined experimental and computational study of ZIF-67, another ZIF made with Co$^{2+}$ as metallic species, has delved into the pre-nucleation species and their transformation mechanisms.\cite{Filez2021} However, the influence of the solvent, temperature and concentration on MOF synthesis mechanisms has not yet been addressed \textit{via} simulation studies up to the moment of this writing.  

In this contribution, we combine classical molecular dynamics (MD) and well-tempered metadynamics (WT-MetaD)\cite{Barducci2008} simulations to study the molecular mechanism of the nucleation of ZIF-8 in DMSO or methanol as a function of the synthesis temperature and the concentration of reactants. The simulations contain explicit solvent and rely on the nb-ZIF-FF force field.\cite{Balestra2022} First, we investigate the formation of early building blocks, including the whole series of ${[\text{Zn}(\text{MIm})_n]}^{2-n}$ complexes ($n$=0-4). We compare the free energy of this process to its analog of binding imidazolate instead of MIm. Subsequently, nucleation is studied \textit{via} a series of simulation setups that mimic syntheses made with different temperatures, solvents and reactant concentrations. We first observe the formation of linear Zn--ligand chains, which then are transformed to incorporate cyclic structures with lifetimes in the order of 1-10 nanoseconds. These pre-nucleation aggregates continue growing until they merge with other aggregates to form a highly connected amorphous phase. This study represents the first computational contribution where the nucleation of a MOF is studied from the reactants to a highly connected, intermediate, state as a function of synthesis conditions. 

This article is organized as follows. Section II includes the methodological details, a detailed decription of all the systems and simulation setups. Section III describes the results obtained and section IV summarizes the conclusions.  

\section{Methods} 

\subsection{System preparation and setup}

To investigate the formation of early building blocks and the nucleation mechanisms of ZIF-8 in DMSO, we conducted MD simulations on two distinct types of systems. The formation of small molecular complexes at the onset of the synthesis process was studied through a system in which a single
$\text{Zn}^{2+}$ ion was coordinated with two $\left( \text{Zn}^{2+} + 2 \, \text{MIm}^{-} \rightleftharpoons [\text{Zn(MIm)}_2] \right)$ or four methylimidazolate ligands  $\left(  [\text{Zn(Im)}_2] + 2 \, \text{MIm}^{-} \rightleftharpoons [\text{Zn(MIm)}_4]^{2-} \right)$. In another series of systems aimed at studying nucleation, two concentration regimes were explored: [$\text{Zn}^{2+}$]= 0.53~M and 2.49~M. These concentrations are on the higher end of those used in direct synthesis experiments (between 0.1 and 0.0001~M);\cite{Manning2023} this is because modeling lower concentrations would imply systems too large to treat from a computational standpoint. The lower concentration system consisted of 96 $\text{Zn}^{2+}$ cations, 192 $\text{MIm}^{-}$ anions, and 2344 DMSO molecules, contained within a cubic simulation box of approximately 69 \AA \ length. This composition led to two systems: one in which the temperature was kept at 298~K (reference system) and another in which it was set to a higher value of T=350~K. The higher concentration one, in turn, was composed of 96 $\text{Zn}^{2+}$ cations, 192 $\text{MIm}^{-}$ anions, and 425 DMSO molecules, inside a cubic box of approximately 43 \AA \ length. Both systems were time-evolved following an unbiased molecular dynamics approach.
All initial configurations were generated using the \textit{Packmol} software\cite{Martnez2009}, ensuring an homogeneous distribution of species while preventing steric overlaps. Charge neutrality was maintained by adjusting the number of $\text{MIm}^{-}$ anions to preserve stoichiometric balance in all cases. Simulations were performed using the \textit{LAMMPS}  software\cite{Thompson2022}, coupled with the \textit{PLUMED} package\cite{Tribello2014} for enhanced sampling in the case of the systems aimed at studying the formation of ${[\text{Zn}(\text{MIm})_n]}^{2-n}$ complexes.

The $\text{Zn}^{2+}$ ions and $\text{MIm}^{-}$ ligands were modeled using the nb-ZIF-FF force field\cite{Balestra2022}, which describes Zn–N interactions through a Morse potential, allowing for the reversible formation and dissociation of coordination bonds. 
This force field also incorporates dummy atoms into both Zn and N atoms to accurately reproduce the tetrahedral coordination environment of Zn$^{2+}$ in a field of negatively charged ligands. All simulations were carried out in the NPT ensemble at T=298,350 K and P=1 bar, with temperature and pressure controlled by Nosé-Hoover thermostats and barostats, with damping times of 100x and 1000x the timestep, respectively. The timestep was set to 0.5 fs. The Ewald method was used to accurately compute long-range electrostatic interactions with a relative error tolerance in the forces of 10$^{-8}$. For van der Waals forces and the real part of the Ewald summation, a cutoff distance of 13 \AA \ was applied. In the WT-MetaD simulations, the multiple walker technique with five walkers was employed to efficiently explore the free energy landscape.\cite{Raiteri2005} Three production runs starting from independent initial configurations were set for each system, covering a total sampling time of 30–150 ns depending on the system studied.

\subsection{Metadynamics}
Metadynamics\cite{Laio2002} is an enhanced sampling technique designed to explore the free energy landscape of complex systems. Briefly, it works by introducing a history-dependent bias potential that gradually fills the energy wells, effectively preventing the system from revisiting previously sampled configurations. This bias is deposited in the space of selected collective variables (CVs) over time, allowing efficient exploration of states that would otherwise be difficult to access.
WT-MetaD\cite{Barducci2008} is an improved version of standard metadynamics, where the height of the Gaussian bias decays over time, ensuring a more controlled and gradual exploration of the free energy landscape. This gradual reduction in bias prevents excessive perturbation while still allowing the system to overcome free energy barriers. As a result, WT-MetaD provides a more accurate reconstruction of the underlying free energy surface.\cite{Barducci2008}
For the system with 1 $\text{Zn}^{2+}$ and 2 $\text{MIm}^{-}$ $\left( [\text{Zn(Im)}_2] \right)$, the CVs were defined as the minimum distance between the $\text{Zn}^{2+}$ cation and the two nitrogen atoms of each ligand. This formulation allows $\text{Zn}^{2+}$ to interact with any of the two available coordination sites on each of the ligands. The minimum distance between the Zn and the two available nitrogen atoms (N$_{1}$ and N$_{2}$) of the ligands, $d^{(i)}_{\text{min}}$, to the $\textit{i}$-th ligand is defined as:
\begin{equation}
d^{(i)}_{\text{min}} = \left( d_{N^{(i)}_1}^{-6} + d_{N^{(i)}_2}^{-6} \right)^{-1/6}, \quad i=1,2
\end{equation}

 The same CVs were used for the system containing  2 $\text{Zn}^{2+}$ with 4 $\text{MIm}^{-}$, but they focus on the two new ligands to be added. The ligands previously attached to the tagged Zn were constrained to stay binded \textit{via} a harmonic restraint. To maintain the electrical neutrality of the system, an extra $\text{Zn}^{2+}$ ion was added, with a constraint to keep it away from the other reactive sites. The formation of the fully coordinated ${[\text{Zn}(\text{MIm})_4]}^{2-}$ complex was carried out in two steps, as using more than three CVs can cause convergence issues in metadynamics (see the supporting information -SI- for metadynamics simulations and convergence details).

\section{Results}
\subsection{Formation of initial building blocks}

\begin{figure*}
\begin{center}
\includegraphics[width=\textwidth]{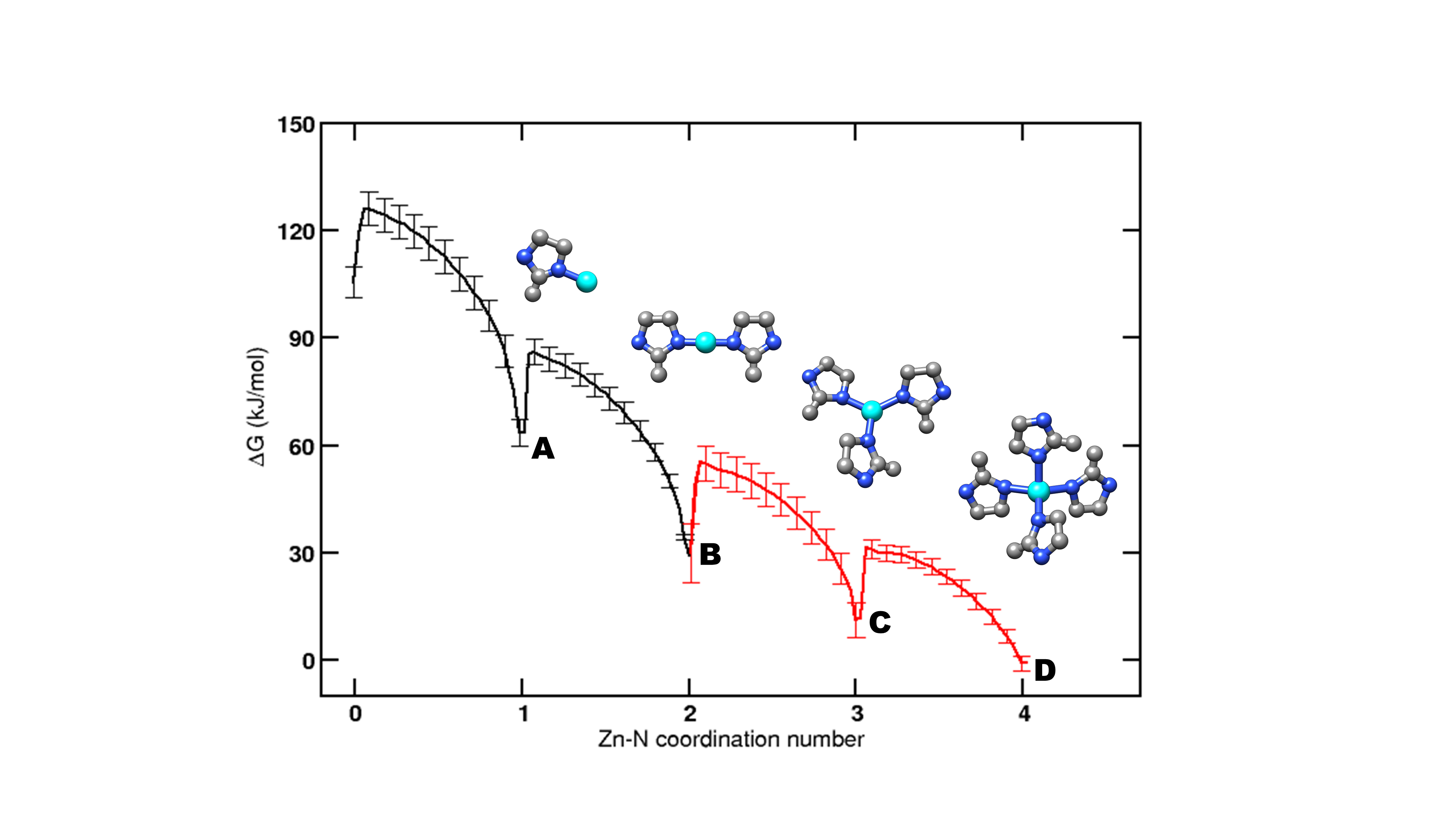}
\caption{ Free energy profile of the sequential binding of ligands to a $\text{Zn}^{2+}$ cation. The black curve represents the pathway leading to the formation of $[\text{Zn(MIm)}_2]$, and the red curve corresponds to the formation of $[\text{Zn(MIm)}_4]^{2-}$. Both curves are aligned at the common intermediate state B. The free energy is set to zero at the global minimum, corresponding to species D, for reference. }
\label{fig:blocks}
\end{center}
\end{figure*}

Fig.~\ref{fig:blocks} shows the free energy profile associated to the sequential binding of ligands to $\text{Zn}^{2+}$ during the self-assembly process. The system starts from a solvated $\text{Zn}^{2+}$ ion, which has a high positive charge and is free to bind with MIm$^-$ ligands. At this point, the free energy is relatively high, as the $\text{Zn}^{2+}$ is surrounded by the neutral solvent molecules. When the first ligand binds to $\text{Zn}^{2+}$, there is a significant drop in free energy $\Delta G_{Zn^{2+}\xrightarrow \ A} = -40 \pm 4\, \text{kJ/mol}$ (state A in Fig.~\ref{fig:blocks}), indicating that this binding event is highly favorable. This large drop can be associated to the strong electrostatic attraction between the negatively charged ligand and the $\text{Zn}^{2+}$. After the first Zn--ligand bond is formed, each additional binding of a ligand still leads to a decrease in free energy, but the drops become progressively smaller $\Delta G_{A\xrightarrow \ B} = -34 \pm 6 \, \text{kJ/mol}$, $\Delta G_{B\xrightarrow \ C} = -18 \pm 7  \, \text{kJ/mol}$, and $\Delta G_{C\xrightarrow \ D} = -11 \pm 3 \, \text{kJ/mol}$ for the addition of the second, third, and fourth $\text{MIm}^{-}$ ligands, respectively (states B, C, and D in Fig.~\ref{fig:blocks}). This indicates that, while each binding event is still favorable, the energetic advantage of binding decreases as more ligands are added. The associated errors are calculated using the method described in the SI.

This result is in contrast with a recent investigation done in our group in which we explored the free energy of formation of ${[\text{Zn}(\text{Im})_n]}^{2-n}$ complexes ($n$=0-4, and Im is imidazolate) in DMF.\cite{Mendez2025} In that case, the solvent--metal/ligand--metal competition was such that the first three ligand additions led to similar drops in free energy, while the last ligand was the only one associated to a lower drop in free energy. Since both solvents have similar sizes\cite{Pacak1987} and the charges of the ions are the same, this difference could be attributed to the size of the ligand, which is larger in this work, leading to greater ligand--ligand repulsions in subsequent additions. This thermodynamic flexibility could be at the onset of the fact that many more polymorphs can be formed in imidazolate based ZIFs than in their 2-methylimidazolate analogs.\cite{Park2006} Indeed, if the free energy change of binding more ligands to the same metal ion is similar (as is the case for the Im-based ZIFs), a larger diversity in pre-nucleation units (more or less ramified) can be formed with similar free energy costs, while for bulkier ligands, linear-like pre-nucleation building blocks are thermodynamically favored over the others. This analysis is complementary to other approaches to studying polymorph stability that are based in the stability of the polymorphs themselves.\cite{Han2018,Rojas2025}

\subsection{Nucleation}

\begin{figure*}[b]
\includegraphics[width=\textwidth]{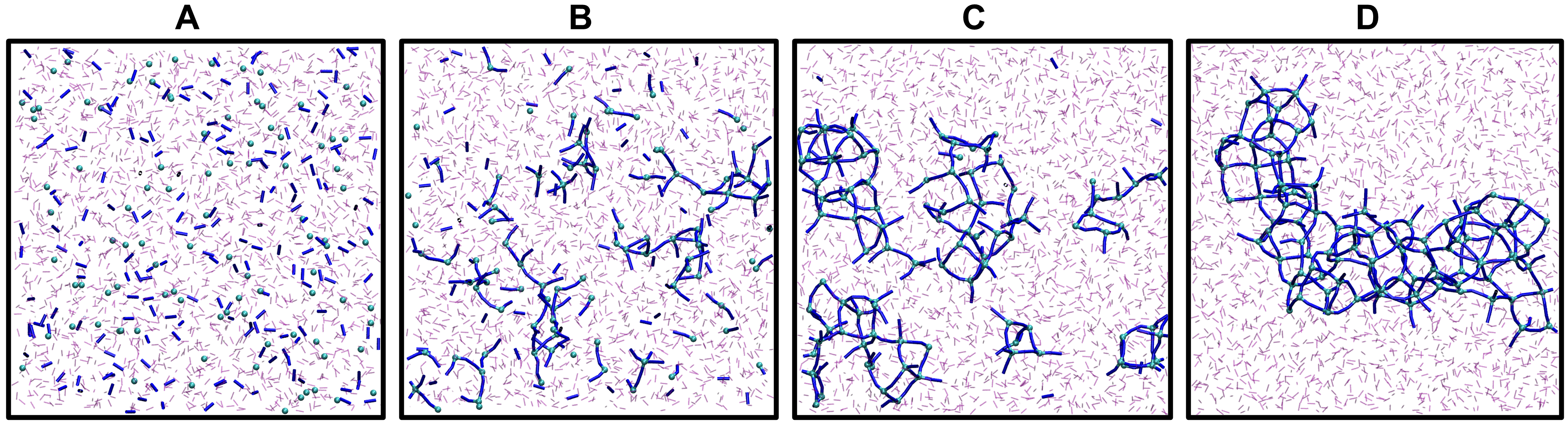}
\caption{ Representative snapshots depicting the structural evolution of the system during the simulation: from separated ions (A) to linear oligomers (B), polymers with rings (C), and an amorphous aggregate (D). $\text{Zn}^{2+}$ ions are shown in cyan spheres, ligands are represented as blue lines that bridge their two nitrogen atoms and solvent molecules are colored in purple and represented with higher transparency. Snapshots were generated using Visual Molecular Dynamics (VMD).\cite{HUMPHREY199633}}
\label{fig:pattern}
\end{figure*}

All nucleation simulations featured some common mechanistic patterns, illustrated in Fig.~\ref{fig:pattern}. First, predominantly linear oligomers of alternating $\text{Zn}^{2+}$ and MIm$^-$ ions were formed. These chains continued to grow and gradually ramify to form cycles or rings. Rings are important structural motifs in porous materials, they are present in the final MOF structure and act as ``windows" for accessing the porosity. At longer simulation times, the oligomers became polymers and merged, to finally yield a highly-connected amorphous phase, that could be associated to that found in direct experiments\cite{Venna2010,Jin2023,Dok2025, Balog2022,Talosig2024} as in previous work.\cite{Balestra2022}

\subsubsection{Time evolution of Zn--ligand coordination number}
 
\begin{figure*}[t]
\includegraphics[width=0.9\textwidth]{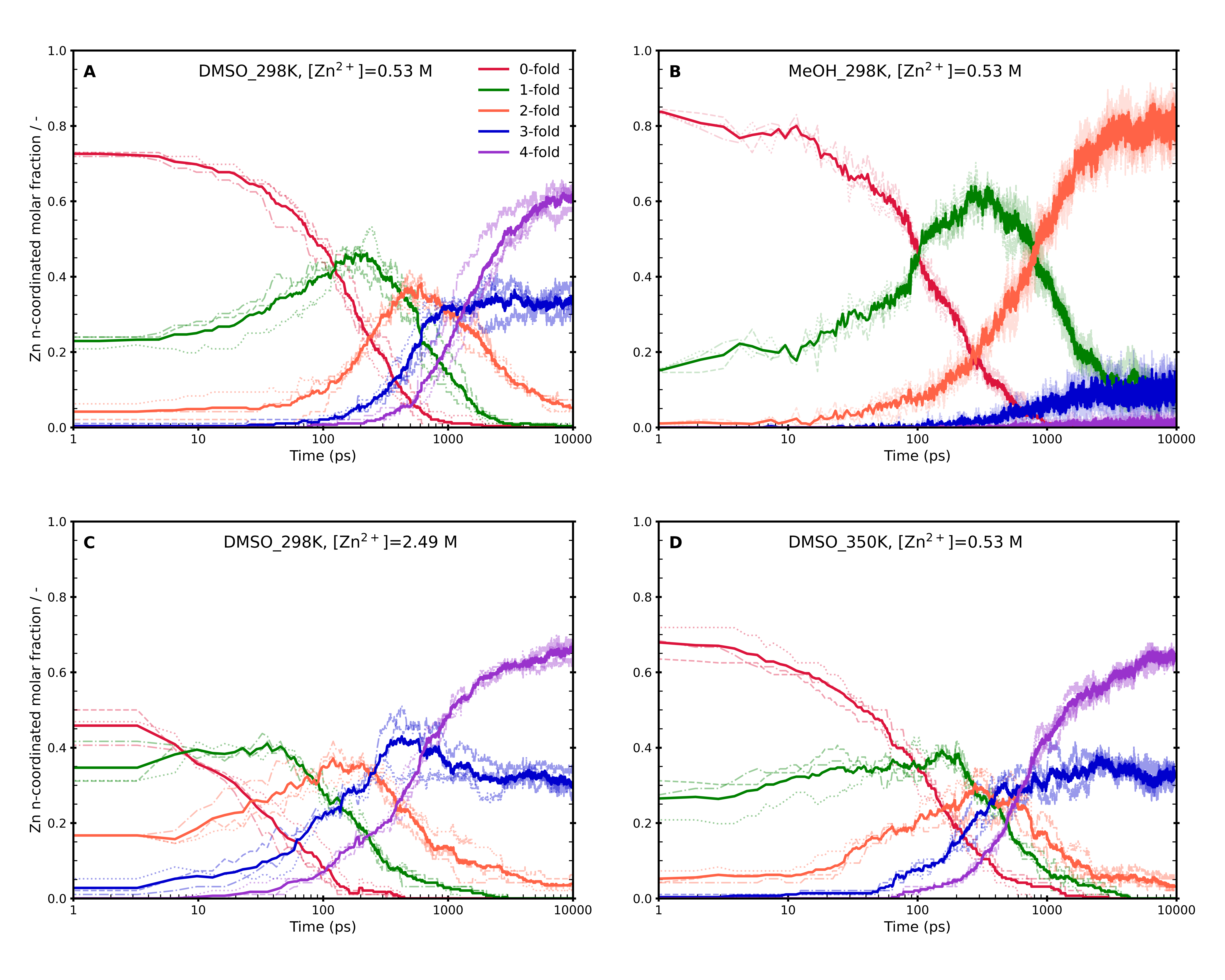}
\caption{ Time evolution of the population of n-fold coordinated $\text{Zn}^{2+}$ cations under different synthesis conditions: in DMSO at 298 K with $\left[\text{Zn}^{2+}\right]$ = 0.53 M (A), in MeOH at 298 K  with $\left[\text{Zn}^{2+}\right]$ = 0.53 M (B), in DMSO at 298 K with $\left[\text{Zn}^{2+}\right]$ = 2.49 M (C), and in DMSO at 350 K with $\left[\text{Zn}^{2+}\right]$ = 0.53 M (D).The average is shown as a opaque highlighted line and independent simulations are represented by transparent dashed, dotted, and dot-dashed lines.}
\label{fig:nucleation}
\end{figure*}

In order to follow the kinetics of the nucleation process, we analyzed the time evolution of the coordination between $\text{Zn}^{2+}$ ions and nitrogen atoms in MIm$^-$ ligands in two different solvents used in real laboratory experiments\cite{Feng2016}: DMSO and MeOH. Simulations were performed at [$\text{Zn}^{2+}$]= 0.53~M and T=298~K.
Fig. ~\ref{fig:nucleation}A and ~\ref{fig:nucleation}B show the time evolution of Zn–-N coordination number for the two solvents. 
In all systems, a significant fraction of $\text{Zn}^{2+}$ ions is initially fully surrounded by solvent molecules and, thus, 0-fold coordinated with respect to the ligands (indicated by the red curves). Despite constraints on bonds formation, a small fraction of 1-fold and 2-fold coordinated $\text{Zn}^{2+}$ is already formed during the equilibration phase.

In DMSO systems (see Fig.~\ref{fig:nucleation}A), the number of 0-fold $\text{Zn}^{2+}$ ions decreases as time progresses, while the populations of 1-fold coordinated and 2-fold coordinated $\text{Zn}^{2+}$ ions increase. As these two latter curves increase, they overlay the 0-fold coordinated curve at approximately 125 ps and 230 ps, respectively, and then the 1-fold coordinated $\text{Zn}^{2+}$ rises until around 300~ps whereas the 2-fold coordinated continues to rise until 700~ps before decreasing gradually. The population of 0-fold coordinated $\text{Zn}^{2+}$ ions drops to zero around 1.9~ns, while the population of 1-fold coordinated $\text{Zn}^{2+}$ ions reaches zero at 7~ns. Meanwhile, the populations of 3-fold and 4-fold coordinated $\text{Zn}^{2+}$ become non-zero and start to increase at 40 and 130 ps, respectively. Both continue to increasee in parallel, mainly at the expense of a reduction in the 0-fold and 1-fold coordinated species, until the 3-fold coordinated curve begins to stabilize at 2~ns. The 4-fold coordinated $\text{Zn}^{2+}$ ion population continues to increase, tending toward stabilization at around 4~ns. 

Similarly, for the MeOH solvent systems (Fig.~\ref{fig:nucleation}B), the population of 0-fold coordinated $\text{Zn}^{2+}$ ions decreases over time, with the curve reaching zero at a similar time around 2~ns, and the 1-fold and 2-fold coordinated $\text{Zn}^{2+}$ ions populations begin to increase and overlap also at similar times than in the DMSO case. However, the nucleation process is much slower in this solvent. The population of 2-fold coordinated $\text{Zn}^{2+}$ ions increases sharply, reaching approximately 85\% of the total amount of $\text{Zn}^{2+}$ ions on average (more than double than what is found for DMSO). After 10~ns of simulation, we find that only about 10\% of the $\text{Zn}^{2+}$ ions are 3-fold coordinated, while the population of 4-coordinated $\text{Zn}^{2+}$ is almost negligible. In contrast, DMSO-containing systems exhibited a notably different trend at the same simulation times, with an average of 68\% of the $\text{Zn}^{2+}$ population being 4-fold coordinated to MIm$^-$ ligands.

The significant difference observed in the kinetics of the nucleation process in DMSO versus MeOH can be correlated with experimental findings. Feng and coworkers found that synthesizing ZIF-8 in DMSO lead to smaller nanoparticles than when the synthesis was carried out in MeOH.\cite{Feng2016} This suggests that nucleation is faster when compared to growth in DMSO. The authors attribute this effect to a slowdown in the growth process in DMSO due to a better solvation of the ionic reactants than in MeOH. Our results suggest that the difference could be already present in the nucleation part of the self-assembly process. Indeed, we observe a slowdown in the diffusion of the reactive ions in methanol solution due to a slower solvent dynamics given by the presence of hydrogen bond networks, in line with the reasonings made by Bustamante and coworkers.\cite{Bustamante2014}

To continue our analysis, we investigate the influence of ion concentration on the ZIF-8 nucleation by comparing systems containing DMSO as a solvent at [$\text{Zn}^{2+}$]= 0.53 M and 2.49 M (see Fig.~\ref{fig:nucleation}A and ~\ref{fig:nucleation}C). For higher concentrations (Fig.~\ref{fig:nucleation}C), the simulation starts with a lower population of 0-coordinated $\text{Zn}^{2+}$ ions compared to the lower concentration case, but with noticeable presence of 1-fold, 2-fold, and 3-fold coordinated species already generated during the equilibration phase. It can be observed that the population of 0-fold coordinated $\text{Zn}^{2+}$ ions decreases rapidly, reaching zero around 350~ps, significantly faster than for the lower concentration case. The 1-fold coordinated $\text{Zn}^{2+}$ population also drops to zero at approximately 2~ns. The 3- and 4-fold coordinated $\text{Zn}^{2+}$ populations increase quickly thereafter, and the 4-fold coordinated fraction increases sharply to 50\% by 1~ns. In comparison, only 20\% of $\text{Zn}^{2+}$ ions had achieved 4-fold coordination at the same simulation time for the lower concentration system. As the simulation progresses, the fully coordinated $\text{Zn}^{2+}$ population in the higher concentration system continues to grow until it eventually plateaus from 2~ns onward. In parallel, the 3-fold and 2-fold coordinated $\text{Zn}^{2+}$ fractions (blue and orange lines) stabilize at around 30\% and 5\%, respectively, and remain steady throughout the simulation time. 

To complete this analysis, we compared the $\text{Zn}^{2+}$--ligand complexes population evolution in DMSO at two different temperatures: 298~K and 350~K (see  Fig.~\ref{fig:nucleation}A and ~\ref{fig:nucleation}D). While both systems show a similar trend, the bond formation was accelerated at the higher temperature, as expected given the faster diffusion of reactants. At 350~K, the average populations of 1-fold and 2-fold coordinated $\text{Zn}^{2+}$ ions never exceed 40\% and 20\%, respectively. In contrast, at 298~K (Fig.~\ref{fig:nucleation}A) the average 1-fold coordinated population rised to 45\% and 2-fold coordinated $\text{Zn}^{2+}$ ions reached 35\%. This means that the $\text{Zn}^{2+}$ ions tend to adopt higher coordinations faster at higher temperature, suggesting that the diffusion of the reactants is one of the slow modes of nucleation. This is supported by the faster increase in the population of 3-fold and 4-fold coordinated $\text{Zn}^{2+}$ ions. After 1~ns, the system at 350~K showed approximately 45\% of $\text{Zn}^{2+}$ ions in the fully 4-fold coordinated state, compared to only about 20\% at 298~K. 

Both increasing temperature and increasing concentration result in faster nucleation kinetics, as expected from entropic considerations. This suggests that metal ion -- ligand encounters are rare events and that their diffusion towards one another consitutes one of the slow degrees of freedom in the nucleation process. The overall shape of the curves for n-fold coordination of Zn$^+$ ions is similar regardless of whether these encounters are favored by an increase in the amount of reactants of by an increase in their kinetic energy (Fig.~\ref{fig:nucleation}C and ~\ref{fig:nucleation}D). In the following sections we will explore the differences that underlie these two synthesis setups.

\subsubsection{Nucleation mechanism}

\begin{figure*}
\includegraphics[width=0.9\textwidth]{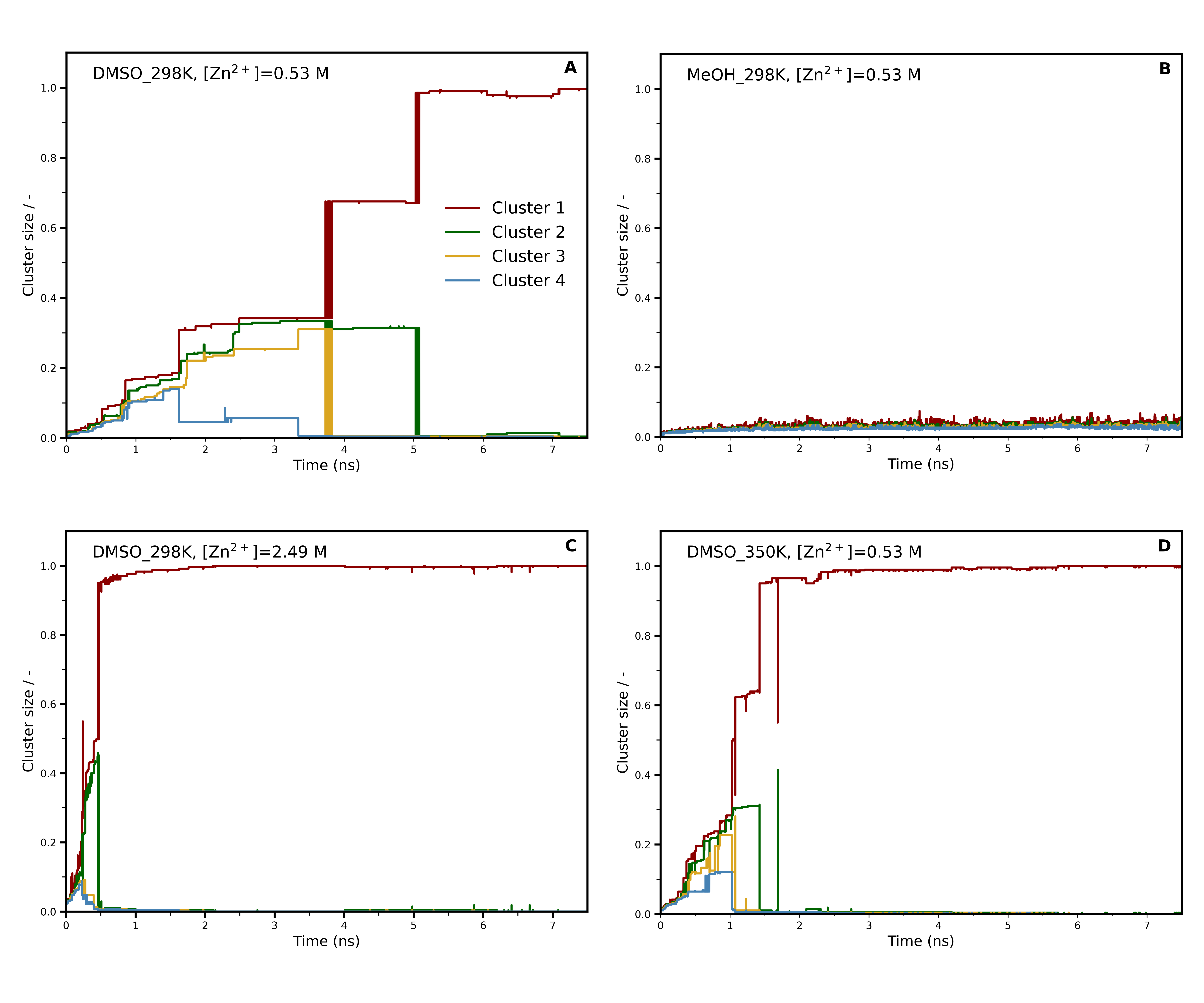}
\caption{ Time evolution of the normalized number of Zn and N atoms in the largest four clusters under different synthesis conditions. In DMSO at 298 K with $\left[\text{Zn}^{2+}\right]$ = 0.53 M (A), in MeOH at 298 K with $\left[\text{Zn}^{2+}\right]$ = 0.53 M (B), in DMSO at 298 K with $\left[\text{Zn}^{2+}\right]$ = 2.49 M (C), and in DMSO at 350 K with $\left[\text{Zn}^{2+}\right]$ = 0.53 M (D).}
\label{fig:cluster}
\end{figure*}

The evolution of nuclei size during the early nucleation stages of ZIF-8 was next examined by analyzing the size of the agreggates or clusters formed by metal ions and ligands that are bonded to each other over time. Specifically, we tracked the growth of the largest four clusters present at all times, as shown in Fig.~\ref{fig:cluster}. The cluster sizes are computed as the number of Zn and N atoms that belong to that cluster, normalized by the total number of these species in the simulation box. A value of 1 therefore corresponds to a single, fully connected cluster including all the ions, while a value of 0 reflects the initial condition of the synthesis process, where the Zn$^{2+}$ and ligands are fully solvated and not connected to each other. Results shown correspond to one of the independent simulations, all other results are included in the SI. 
Initially, each system features the formation of several clusters, which increase in size in a parallel fashion. These early clusters gradually merge with one another, leading to the formation of larger structures. Nuclei growth mainly occurs by the merging between aggregates, with little evidence of metal or ligand dissociation from existing aggregates. This behaviour is different to what was previously observed for the ZIF-8 nucleation in MeOH\cite{Balestra2022}, where clusters could merge but could also dissolve to replenish the free ions pool. Fig. ~\ref{fig:cluster}A shows that at around 1.6 ns, cluster 4 merges with cluster 1, resulting in a noticeable increase in the size of cluster 1. Meanwhile, clusters 2 and 3 continue to grow in parallel. Around 3.5 ns, cluster 4 further merges with cluster 3, and after that, clusters 1 and 2 are stable for 1 ns, until when cluster 2 joins cluster 1, forming one large cluster at 5 ns. A movie illustrating this cluster evolution is provided as SI.

By changing the solvent to MeOH (Fig ~\ref{fig:cluster}B), it is observed that until about 7 ns of simulation, all clusters have similar, small sizes. This clearly indicates that nucleation is considerably slower in this solvent. In contrast, Fig. ~\ref{fig:cluster}C shows that increasing the concentration of reactants leads to significantly faster nucleation. Two out of the three independent simulations show an aggregate merging mechanism, as that found for the reference system. The third one (see SI) exhibits a single, large cluster that is formed early in the simulation. This is possible given the stochastic character of the process and that the concentration is very high in this system.

In what concerns temperature, Fig. ~\ref{fig:cluster}D shows a similar trend as that observed for its analog at ambient temperature in Fig. ~\ref{fig:cluster}A, but with faster cluster growth due to the higher kinetic energy that contributes to faster diffusion of the species. At around 1 ns, clusters 3 and 4 attach to cluster 1, followed by cluster 2 joining cluster 1 at approximately 1.5 ns. This process is completed within about 2 ns, after which only one cluster persists in the system. Additional independent simulations were performed for each system to verify reproducibility. The results show consistent cluster growth trends and are presented in the SI.

\subsubsection{Formation of rings}

In this section, we study the structure and dynamics of ring formation under different synthesis conditions, and we will unveil differences between the amorphous structures formed either by increasing concentration of reactants or by increasing temperature. Rings are typically referred to as n-membered, here we will adopt the convention of setting n equal to the number of Zn$^{2+}$ ions that compose them. Even though ZIF-8 only contains 4- and 6-membered rings, we have observed that 3-, 5-, 7- and 8- membered rings are also formed, which has also been reported in previous work where amorphous ZIFs were studied.\cite{Castel2024} This is reasonable if we think that the synthesis of ZIF-8 proceeds via a highly-connected amorphous intermediate species.\cite{Venna2010,Dok2025,Talosig2024}

\begin{figure*}[t]
\includegraphics[width=0.78\linewidth]{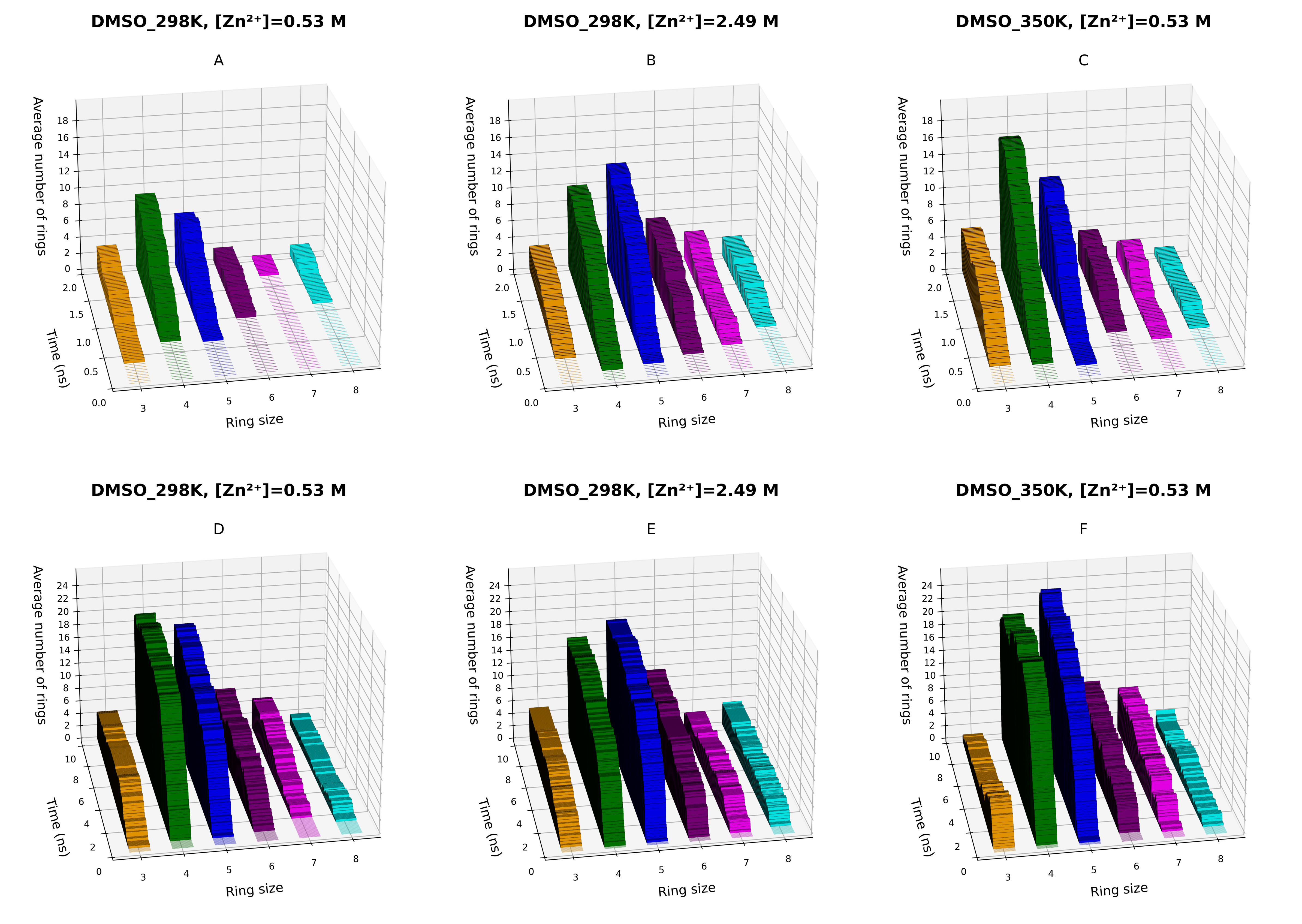}
\caption{ Evolution of the average number of rings of different sizes as a function of simulation time for the three systems containing DMSO as a solvent: T=298~K and $\left[\text{Zn}^{2+}\right]$ = 0.53~M (A,D), T=298~K and $\left[\text{Zn}^{2+}\right]$ = 2.49 M (B,E), and T=350~K and $\left[\text{Zn}^{2+}\right]$ = 0.53~M (C,F). Ring counts are averages from three independent simulations in each case. Plots labeled A, B and C display data from the initial 2~ns to emphasize the early stages of ring formation, whereas plots labeled D, E and F present results up to 10~ns, to illustrate how ring populations stabilize at longer timescales. }
\label{fig:ringsize}
\end{figure*}

Fig.~\ref{fig:ringsize} presents the evolution of the number of rings of different sizes as a function of simulation time for the three DMSO-containing synthesis systems, averaged over three independent simulations in each case. Plots labeled A, B and C focus on the first 2~ns, providing detailed information on the initial dynamics of ring formation. In contrast, plots labeled D, E and F show the longer-term evolution of ring populations. 
The evolution of ring populations over the first 2~ns of simulation time across all systems (Fig.~\ref{fig:ringsize}A, ~\ref{fig:ringsize}B and ~\ref{fig:ringsize}C) reveals that the sequence of ring formation changes as a function of synthesis conditions. The earliest ring formations differ across systems, with 3-, 4- and 5-membered rings appearing first in the reference, higher concentration and higher temperature system respectively. The 7- and 8-membered rings take the longest to appear as expected, since they require for more coordination bonds to form. The average number of 4- and 5-membered rings continues to grow in all systems. 4-membered rings are predominant in the reference and higher T systems, while 5-membered rings dominate the higher concentration case. 

At longer timescales (see Fig. ~\ref{fig:ringsize}D, ~\ref{fig:ringsize}E and ~\ref{fig:ringsize}F), we see a stepwise increase in the average number of 3-membered rings for the ambient temperature systems, which eventually stabilizes toward the end. In contrast, the higher temperature system shows an initial increase in the average number of 3-membered rings, followed by a noticeable decrease. Conversely, 4-, 5- and 6- membered rings feature a continuous increase and 7- and 8-membered rings quickly reach a plateau in all cases.

It is possible to see that the final ring distribution differs according to the system, suggesting a difference in the highly-connected, amorphous structures that are formed upon nucleation occuring under different synthesis conditions. Only the reference system features 4-membered rings as the most abundant species, while the other two cases for which the kinetics of the process is accelerated show 5-membered rings as the predominant species instead. Increasing concentration seems to promote the formation of 6-membered rings more than increasing temperature. However, this also increases the population of 3-membered rings, which will need to be converted into 4- or 6-membered rings at later synthesis stages to yield a non-defective ZIF-8 material. These results suggest that increasing the temperature could provide the energy needed for the system to overcome the barrier required to eliminate 3-membered rings.

To finalize our analysis, we explored ring lifetimes. For this, we implemented a filtering protocol to ensure that short-timescale distortions in ring structures caused by thermal fluctuations would not be considered as a ring breaking event. Otherwise, rings that clearly were structurally stable by visual inspection would have been recorded as multiple short-lived events, which would lead to the underestimation of their lifetimes. This is a consequence of using strict definitions to define whether a system is connected or not (a fixed distance cutoff of 3 \AA \ between Zn and N). Thus, after plotting the lifetimes and noticing that a bimodal distribution was obtained, we decided to discard rings with lifetimes shorter than 200 ps from the analysis, which covers for the characteristic timescale of rapid thermal fluctuations in Zn--N bonds that generates the first peak in the distribution.

\begin{figure*}[t]
\includegraphics[width=0.69\linewidth]{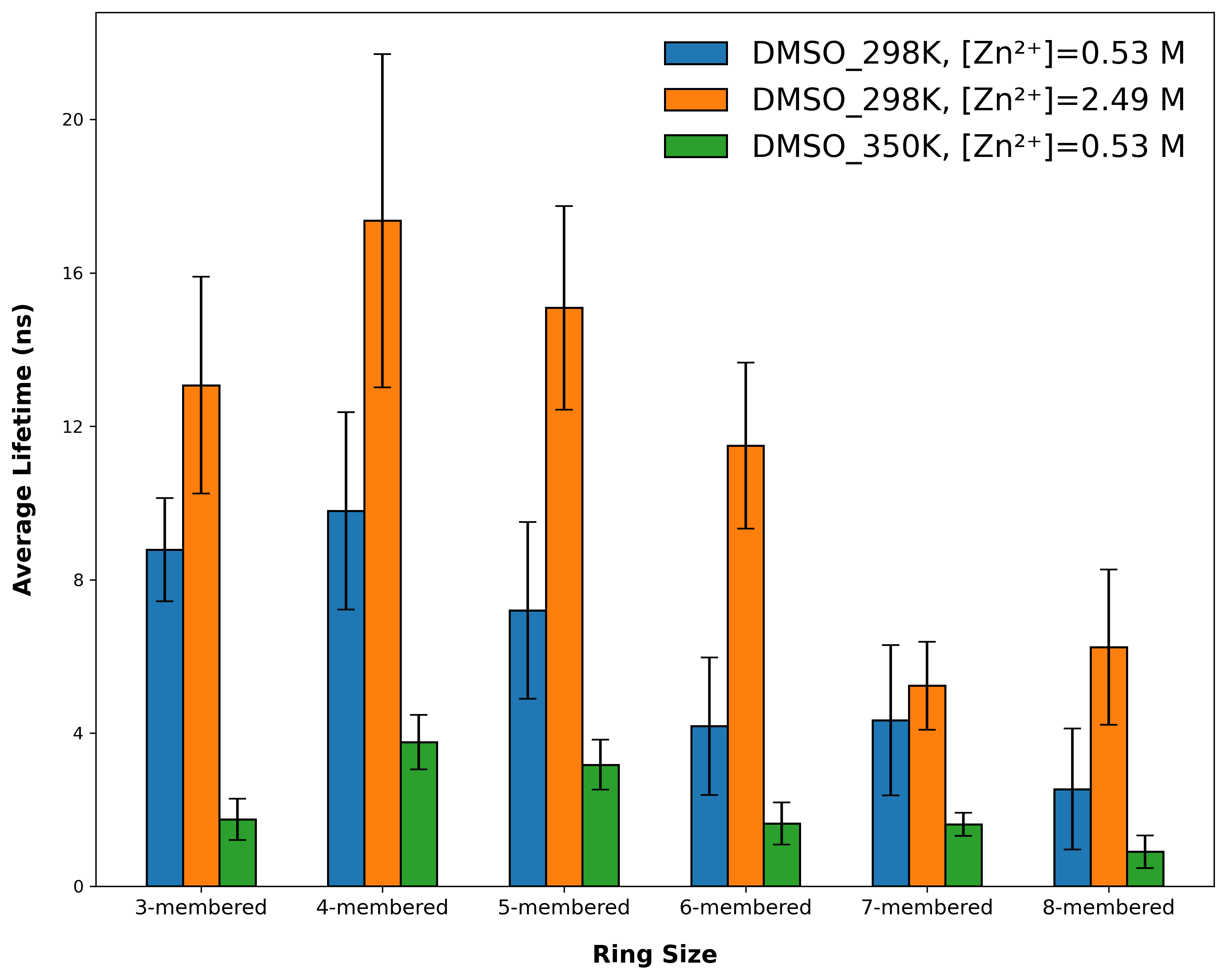}
\caption{ Average lifetimes of n-membered rings (n = 3–8) observed in the three systems containing DMSO as a solvent: T=298~K and $\left[\text{Zn}^{2+}\right]$ = 0.53~M (blue), T=298~K and $\left[\text{Zn}^{2+}\right]$ = 2.49 M (orange), and T=350~K and $\left[\text{Zn}^{2+}\right]$ = 0.53~M (green). Bars represent the average values obtained from three independent simulations for each system, with error bars denoting the standard deviation among those values.}
\label{fig:lifetime}
\end{figure*}

Fig.~\ref{fig:lifetime} shows the average lifetimes of n-membered rings (n=3-8) across all DMSO-containing systems. Each bar represents the mean of the average lifetimes calculated for each independent simulation of the system, with error bars showing the standard deviation across them. 
The longest average lifetimes are those corresponding to the higher concentration case (orange bar), followed by the reference system (blue bar), and finally the higher temperature one (green bar). This trend holds regardless of the ring size, and constitutes another significant difference observed between the cases where concentration or temperature are increased. Indeed, our results suggest that increasing synthesis temperature leads to an amorphous intermediate that is more dynamic, with rings with lower lifetimes, than that generated by increasing reactant concentration. Among the different rings, the 4-membered ones exhibit the highest average lifetime across all three systems, which is reasonable when considering their abundance in the ZIF-8 topology (see Table 1 for all values). In contrast, the shortest-lived rings differ depending on the system: the minimum average lifetime corresponds to the 8-membered ring in the systems with higher concentration or temperature, while the 7-membered ring exhibits the lowest stability in the reference system.
Although 6-membered rings are also abundant in ZIF-8, the data reveal that these rings have lower lifetimes than the 4-membered rings. This observation is consistent with previous reports\cite{Balestra2022} indicating that 6-membered rings have shorter lifetimes due to their higher dynamic nature.

We further analyzed the ring dynamics across all DMSO-containing systems  only considering times for which the amorphous structure was already formed, which resulted in  minimal changes in average ring lifetimes (see SI). For both the higher temperature and reactants concentrations systems, the lifetimes remain similar because the network is stabilized very fast, as large aggregates form after 1-3 ns. However, there are small but noticeable differences for the reference system particularly for 3-, 4- and 8-membered rings: the average lifetime decreased for 3- and 4-membered rings because they form early during amorphization before the fully aggregate structure forms and remain stable, while the 8-membered rings exhibit longer lifetimes but with higher standard deviations due to fewer occurrences and greater variability in their persistence. A complete summary of the average lifetimes calculated for each independent simulation performed is presented in the SI, offering further insight into the reproducibility and variability within the results due to the stochastic character of the process.

\begin{table}[htbp]
\begin{center}
\tiny
\begin{tabular}{ |P{0.3
cm}|P{4.5cm}|P{4.5cm}|P{4.5cm}|  }
 \hline
\textbf{n} & \textbf{T=298\,K, $\left[\text{Zn}^{2+}\right]$ = 0.53\,M} & \textbf{T=298\,K, $\left[\text{Zn}^{2+}\right]$ = 2.49 M} & \textbf{T=350\,K, $\left[\text{Zn}^{2+}\right]$ = 0.53~M} \\
\hline
3 & 9 \ $\pm$ 1~ns & 13 $\pm$ 3~ns & 1.8 $\pm$  0.5~ns \\
\hline
4 & 10 \ $\pm$ 3~ns & 17 $\pm$ 4~ns & 3.8 $\pm$ 0.7~ns \\
\hline
5 & 7 \ $\pm$ 2~ns & 15 $\pm$ 3~ns & 3.2 $\pm$ 0.7~ns \\
\hline
6 & 4 \  $\pm$ 2~ns & 12 $\pm$ 2~ns & 1.6 $\pm$ 0.6~ns \\
\hline
7 & 4  \ $\pm$ 2~ns & 5 $\pm$ 1~ns & 1.6 $\pm$ 0.3~ns \\
\hline
8 & 3 $\pm$ 2~ns & 6 $\pm$ 2~ns & 0.9  $\pm$ 0.4~ns \\
\hline
\end{tabular}
\caption{Average lifetimes of ring sizes (n = 3–8) across the three DMSO-containing systems. Values represent mean lifetimes from three independent simulations per system, with error bars showing standard deviation.}
\end{center}
\end{table}

\section{Conclusions} 

In this work, we have modeled the early stages of the synthesis process of ZIF-8 by molecular simulations relying on a force field that captures the metal--ligand bond formation/breaking dynamics.  

Well-tempered metadynamics simulations were performed to study the free energy surface of the formation of building blocks consisting of Zn--2-methylimidazolate complexes. We found that every new ligand addition involves a lower drop in free energy, which could be due to a combination of charge and steric effects. When comparing the free energy surface of this process to the one associated to its analog for the binding of smaller imidazolate ligands,\cite{Mendez2025} important differences can be unveiled. In this case, the first three out of four ligand additions are thermodynamically equally favorable, which allows for a greater variety of pre-nucleation species, and could be at the onset of the existence of multiple polymorphs for imidazolate based ZIFs.

Our unbiased molecular dynamics simulations exhibited an overall nucleation mechanism consisting of the formation of linear-like oligomers that gradually incorporate cycles of different sizes and lifetimes in the nanosecond realm. These oligomers turned into polymers in time, which then formed aggregates that bind together to generate a highly-connected amorphous phase.

Nucleation was found to be faster in DMSO than in methanol, which can be correlated to the fact that significantly smaller ZIF-8 nanoparticles can be experimentally obtained in DMSO. Increasing either the temperature or the concentration of the reactants leads to faster nucleation kinetics, as expected from entropic considerations. The diffusion of the ionic species that lead to succesful encounters are rare events, and constitute some of the slowest modes in the nucleation process. Even though the acceleration in the nucleation obtained by increasing temperature or reactant concentration is comparable, the structure and dynamics of the highly-connected amorphous phase formed in both cases differs. Indeed, higher temperatures lead to more dynamic structures with lower lifetimes for all rings, which could be helpful preparing the system for the formation of proto-crystals containing only 4- and 6-membered rings. Increasing concentration, in turn, leads to higher proportions of 6-membered rings, but also leads to the creation of more 3-membered rings, which will then need to disappear to yield the final ZIF-8 material. Our results suggest that high temperatures may favor the dissolution of the 3-membered rings formed in the early stages of nucleation while increasing concentration may help form 6-membered rings.    

Our simulations have allowed us to explore the part of the self-assembly process that goes from the formation of early complexes that constitute building blocks, up to the formation of a highly-connected intermediate amorphous species. Between these states, an important part of the nucleation process takes place, and many insights can be obtained. Beyond this point, we expect to observe structural reorganizations including further ring population changes and ordering. Modeling these self-assembly stages is still out of reach, but will be the object of further work in our group. 

\section{Acknowledgments}
This work was funded by the European Union ERC Starting grant MAGNIFY, grant number 101042514. This work was granted access to the HPC resources of IDRIS under the allocations A0150911989 and A0170915688 made by GENCI.

\section{Data availability}

The data supporting this article have been either included as part of the Supporting Information (metadynamics convergence criterion and error bars determination, additional results, movie captions) or uploaded in github at \url{https://github.com/rosemino/MAGNIFY/tree/main/ZIF-8_nucleation_T_concentration_solvent} (force field files and metadynamics parameters).

\section{Supporting Information}

\subsection{Metadynamics}

\subsubsection{Free energy surface}

Fig. 7 represents the two-dimensional free energy surfaces as a function of the minimum distance between $\text{Zn}^{2+}$ and the nitrogen atoms of two 2-methylimidazolate $\text{(MIm}^{-}$) ligands: (i) for a system containing one $\text{Zn}^{2+}$ and two ligands, and (ii) for a system with two $\text{Zn}^{2+}$ and four ligands. The color scale represents free energy, where darker regions indicate lower energy (more stable states). In both panels, clear free energy basins are visible. In (i), the most stable basin corresponds to a $\text{Zn}^{2+}$ coordinated by two ligands, while in (ii), the deeper minimum reflects the formation of the tetra-coordinated complex. In panel (ii), the system started with two ligands already bonded to $\text{Zn}^{2+}$, which were restrained using harmonic potentials to preserve their coordination. The collective variables were defined to monitor the binding of two additional ligands by measuring the minimum distances between $\text{Zn}^{2+}$ and the new ligands. An extra $\text{Zn}^{2+}$ ion was added to ensure charge neutrality and a distance restraint was applied to prevent it to bind with the ligands.

\begin{figure*}[b]
\vspace{-0.5cm}
\includegraphics[width=0.7\textwidth]{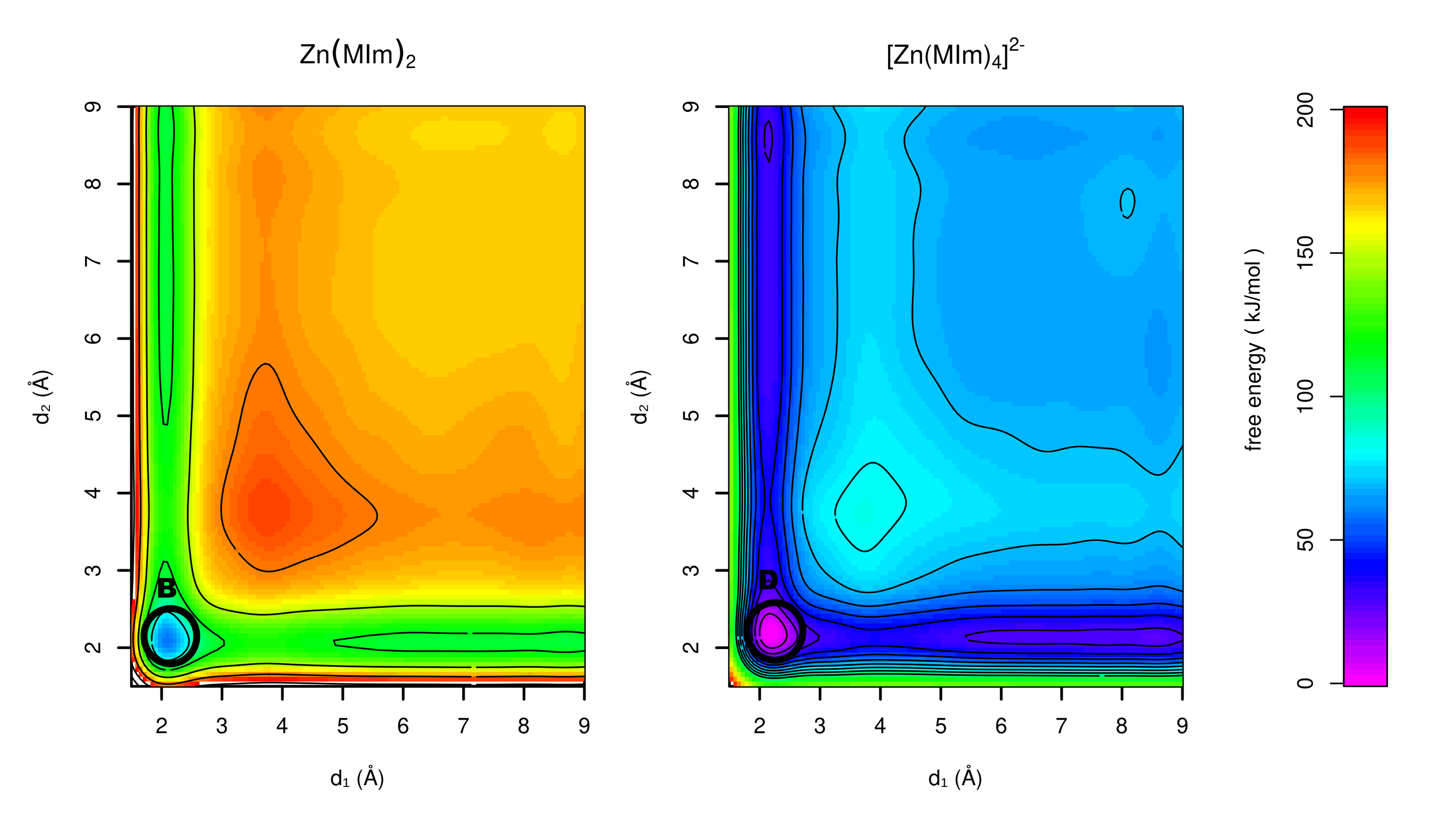}
\vspace{-1cm}
\caption{\label{fig:wide} Two-dimensional free energy surfaces as a function of the minimum distances between $\text{Zn}^{2+}$  and the nitrogen atoms of the ligands. In the left panel, the minimum corresponds to $\text{Zn}^{2+}$ coordinated with two ligands (state B, see Fig. 1 main text), with the light blue regions represent a $\text{Zn}^{2+}$  bonded to only one ligand. In the right panel, the deepest minimum corresponds to a fully coordinated $\text{Zn}^{2+}$ (four ligands, state D see Fig. 1 main text), while the dark blue regions indicate a $\text{Zn}^{2+}$  coordinated with three ligands. }
\end{figure*}

\subsubsection{Convergence of the WT-MetaD simulations}

To assess the convergence of our WT-MetaD simulations, we employed the methodology introduced by Tiwary et al.\cite{Tiwary2014} This approach accounts for the continuous evolution of the bias potential, which prevents it from reaching a fixed plateau value, complicating the definition of a convergence criterion. Tiwary and colleagues tackled this challenge by deriving a time-independent free energy estimator, allowing for a direct comparison of results sampled at different simulation times.

\begin{equation}
G(s) = -\frac{\gamma V(s,t)}{\gamma- 1}  + k_b T \ln \left( \int {\frac{ \gamma V(s,t)} {e^{(\gamma- 1) k_b T}}} \, ds \right)
\end{equation}

Here, s denotes the collective variable(s), ${\gamma}$ is the bias factor, and V(s,t) represents the time-dependent bias potential. The integral in Eq. (2) acts as a time-dependent constant, ensuring consistency between free energy estimates obtained at different time intervals. To extend this analysis to multi-walker simulations, we chronologically ordered the Gaussians from each simulation.
Following this approach, we checked convergence by evaluating how the free energy estimator evolved over time (Fig. 8). However, for our analysis, we specifically selected three points from the 2D surfaces shown in Fig. 7. For the $\text{Zn(MIm)}_2$ system, the selected points are (i) the global minimum, where $\text{Zn}^{2+}$ is bonded to two ligands represented in the 2D free energy plot at a distance of $\text{d}_{1}$ = 2 \AA, $\text{d}_{2}$ = 2 \AA, (ii) a local minimum in which $\text{Zn}^{2+}$ is bonded to a single ligand, with $\text{d}_{1}$ = 2 \AA \ (one ligand bound) and $\text{d}_{2}$ = 6 \AA \ (the second ligand far away), and (iii) same as (ii), but with the order of the ligands inversed ($\text{d}_{1}$= 6 \AA, $\text{d}_{2}$= 2 \AA).
In the case of the ${[\text{Zn}(\text{MIm})_4]}^{2-}$ system, the selected points are: the global minimum, that corresponds to the fully coordinated complex with four ligands bound to $\text{Zn}^{2+}$, and two distinct local minima, each with $\text{Zn}^{2+}$ bound to three ligands ( $\text{d}_{1}$ = 2 \AA,  $\text{d}_{2}$ = 6 \AA \ and $\text{d}_{1}$ = 6 \AA,  $\text{d}_{2}$ = 2 \AA). By choosing these specific points, we ensured that our analysis captures both the most stable structures and the relevant intermediate binding states. 

\begin{figure*}
\includegraphics[width=0.8\textwidth]{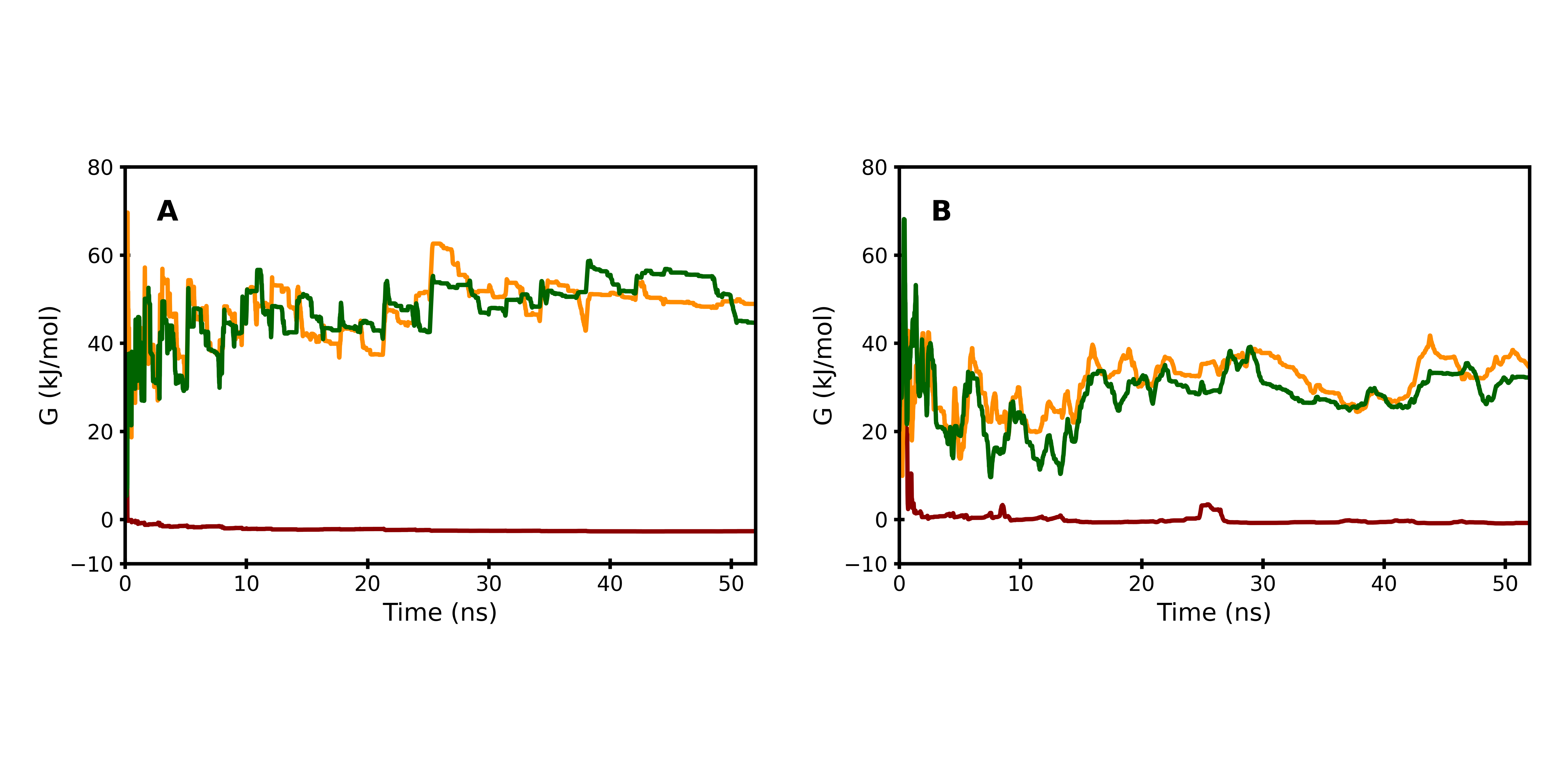}
\vspace{-1.5cm}
\caption{\label{fig:wide} Free energy estimator for three representative points on the 2D free energy surface. Panel A corresponds to the $[\text{Zn(MIm)}_2]$ system and Panel B to the ${[\text{Zn}(\text{MIm})_4]}^{2-}$ system. The red curves in systems A and B represent the global minimum in the free energy landscape and the yellow and green curves correspond to two distinct local minima, where each $\text{Zn}^{2+}$ is coordinated to one or three ligands, respectively.}
\end{figure*}

\subsubsection{Uncertainty calculation}

To reliably determine the final free energy profile and its associated errors from the free energy curves computed with equation (1), time-averaging of these results is essential. However, since simulation data can be correlated, which might introduce inaccuracies, we employed the block averaging technique developed by Bussi and Tribello. \cite{Bussi2019} This method is crucial for estimating the optimal block size where the data becomes uncorrelated for the averaging. This optimal size is identified by plotting the standard deviation of the free energy against the block size. When the standard deviation reaches a plateau, it means that the individual block values are decorrelated (see Fig. 9). To ensure a robust estimation of errors, we averaged data from blocks of  38 and 45~ns for the formations of $[\text{Zn(MIm)}_2]$ and ${[\text{Zn}(\text{MIm})_4]}^{2-}$, respectively.

\begin{figure*}[b]
\includegraphics[width=0.6\textwidth]{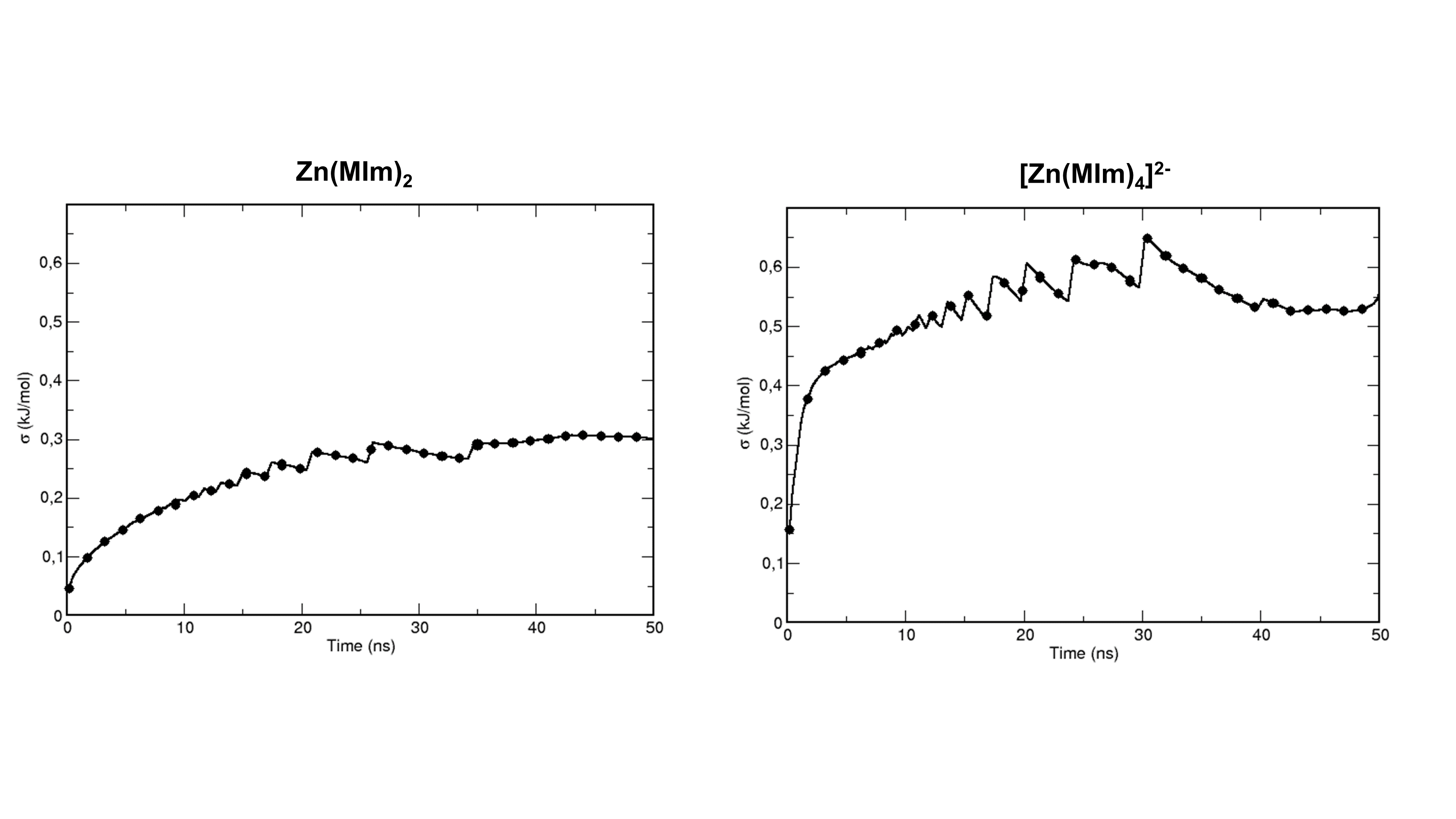}
\vspace{-1.8cm}
\caption{ The figure illustrates the standard deviation of the free energy associated with the lowest energy structure as a function of block size for both $[\text{Zn(MIm)}_2]$ and ${[\text{Zn}(\text{MIm})_4]}^{2-}$ systems, which is crucial for identifying when the simulation data becomes uncorrelated.}
\label{fig:error_block}
\end{figure*}

\subsection{Additional Results} 

\subsubsection{Nucleation analysis}

\begin{figure*}[b]
\vspace{-3cm}
\includegraphics[width=0.65\textwidth]{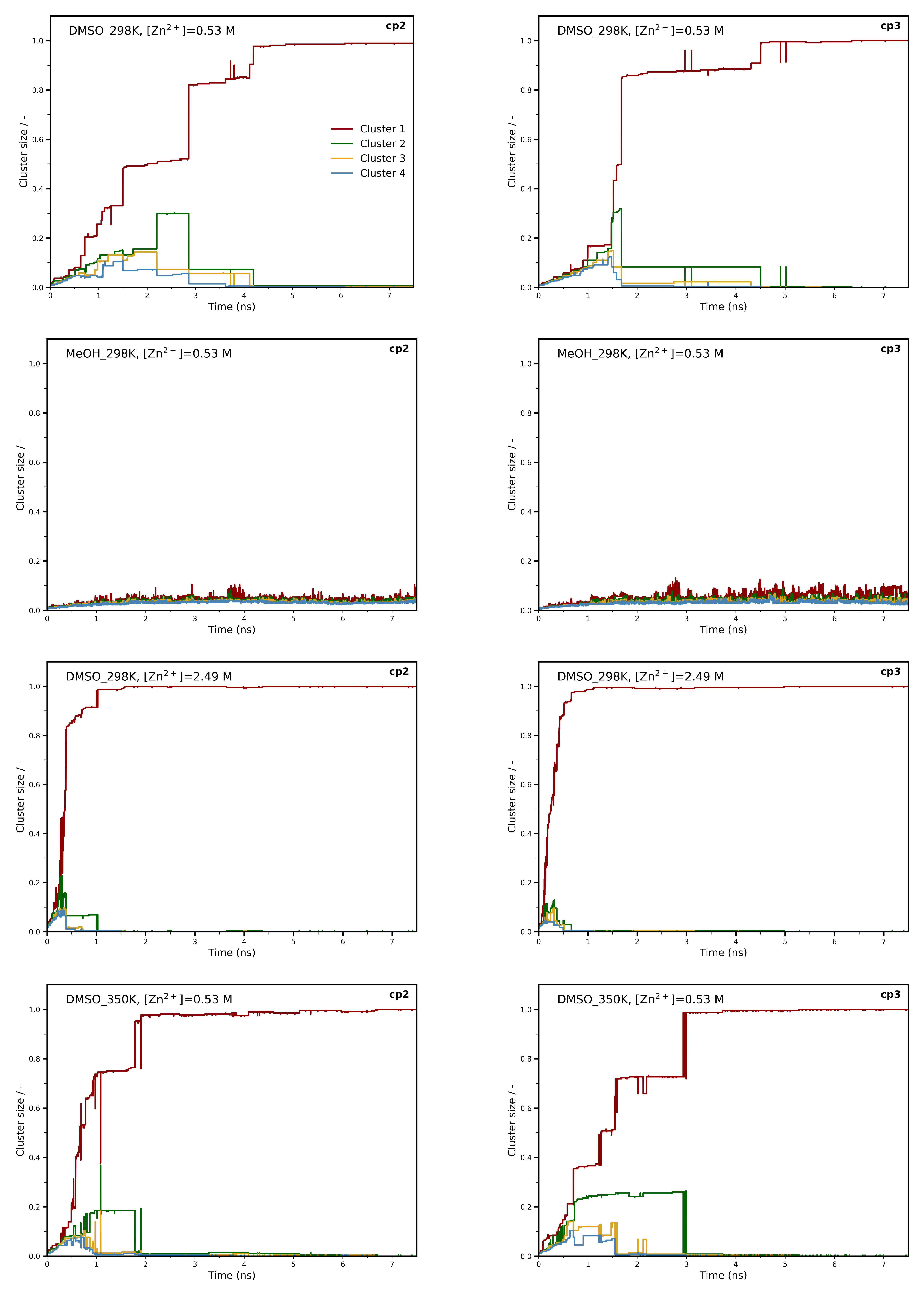}
\caption{ Time evolution of the proportion of Zn and N atoms present in the largest four clusters for four different systems: DMSO at 298~K with $\left[\text{Zn}^{2+}\right]$ = 0.53 M, MeOH at 298~K with $\left[\text{Zn}^{2+}\right]$ = 0.53 M, DMSO at 298~K with $\left[\text{Zn}^{2+}\right]$ = 2.49~M, and DMSO at 350~K with $\left[\text{Zn}^{2+}\right]$ = 0.53 M. The second (left) and third (right) independent simulations of each system are displayed. Each curve represents how the clusters grow over time, providing insight into nucleation dynamics under varying synthesis conditions. Red, green, yellow and light blue curves correspond to the first, second, third and fourth largest cluster, respectively.}
\label{fig:cluster}
\end{figure*}

\begin{figure*}
\includegraphics[width=\textwidth]{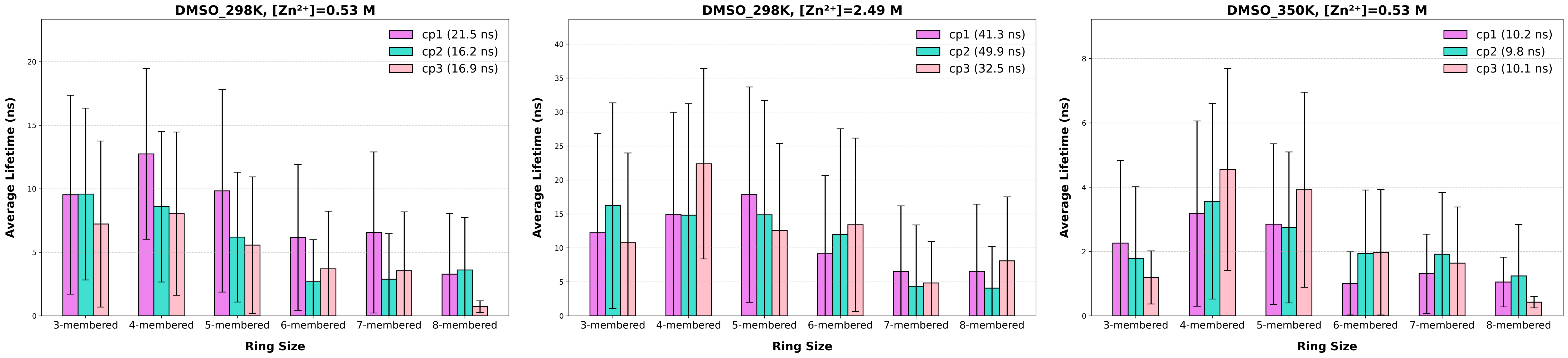}
\caption{ Average lifetimes of n-membered rings (n = 3–8) present in the three DMSO-containing systems: T=298~K with $\left[\text{Zn}^{2+}\right]$ = 0.53~M, T=298~K with $\left[\text{Zn}^{2+}\right]$ = 2.49 M, and T=350~K with $\left[\text{Zn}^{2+}\right]$ = 0.53~M. Bars represent the average lifetime for each independent simulation, and the standard deviation indicates the dispersion of lifetimes within each of them.}
\label{fig:lifetimeSI}
\end{figure*}

\begin{figure*}[t]
\includegraphics[width=0.6\linewidth]{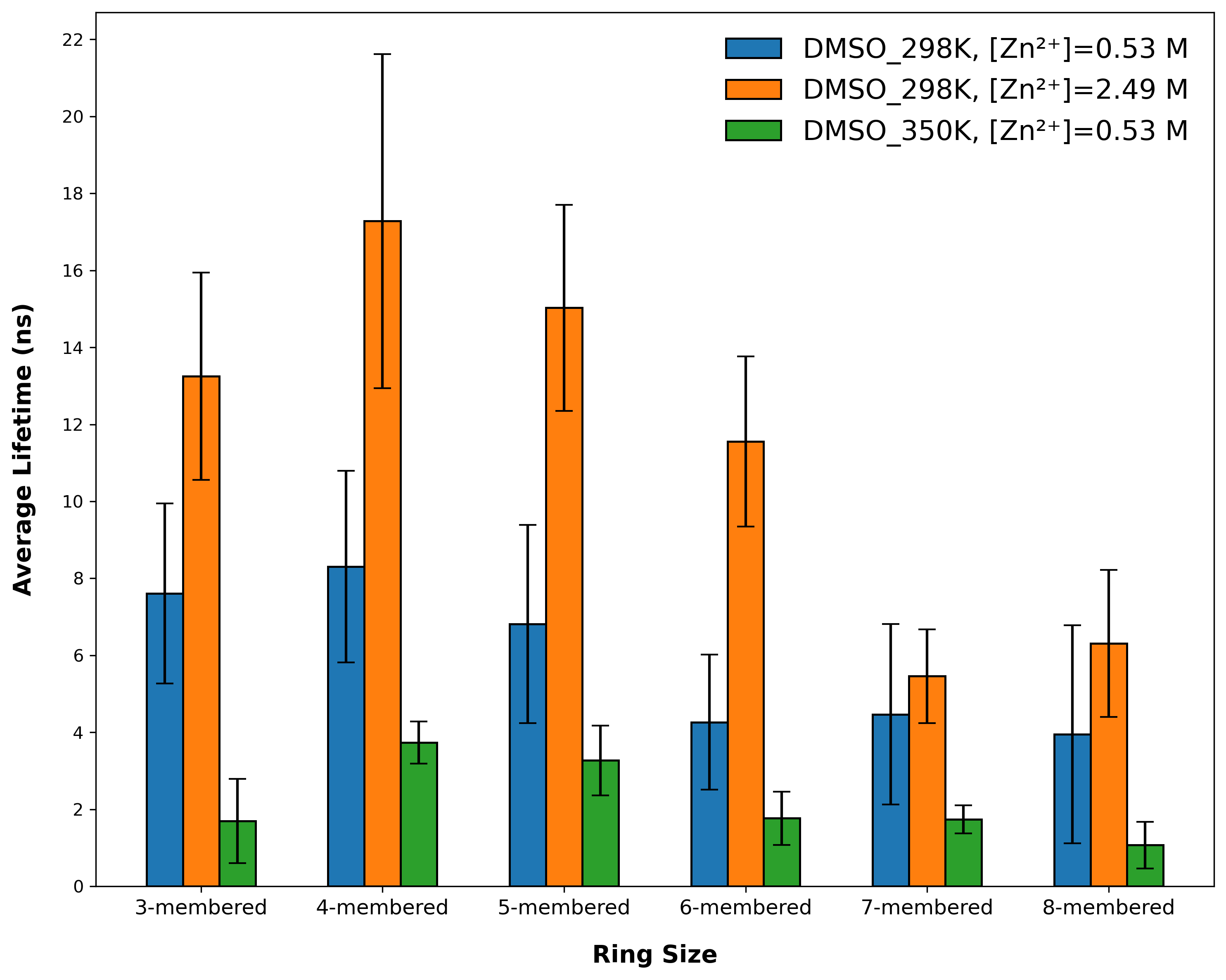}
\caption{Average lifetimes of n-membered rings (n = 3–8) measured in fully amorphous DMSO-solvated systems: T=298~K and $\left[\text{Zn}^{2+}\right]$ = 0.53~M (blue), T=298~K and $\left[\text{Zn}^{2+}\right]$ = 2.49 M (orange), and T=350~K and $\left[\text{Zn}^{2+}\right]$ = 0.53~M (green). Bars show average ring lifetimes from three independent simulations after amorphous aggregate formation, with error bars denoting the standard deviation among them.}
\label{fig:amourphSI}
\end{figure*}

\newpage

\subsubsection{Movie details} 

Comparison of cluster formation dynamics between systems: A) DMSO at 298~K with $\left[\text{Zn}^{2+}\right]$ = 0.53~M (independent simulation 1, labeled cp1) and B) DMSO at 298~K with $\left[\text{Zn}^{2+}\right]$ = 2.49~M (independent simulation 3, labeled cp3). Some intermediate frames were omitted to concisely demonstrate cluster growth processes while reducing movie length and file size. The colors of the four largest clusters match those used in Fig. 4 and Fig. 10. 

\bibliography{magnify}

\end{document}